# The Role of Generative AI in Global Diplomatic Practices
# A Strategic Framework

Muneera Bano[1], Zahid Chaudhri[2], Didar Zowghi[3]

[1] CSIRO's Data61, Clayton, Melbourne, Australia
muneera.bano@csiro.au

[2] High Commission for Pakistan, Canberra, Australia
zchaudhri@gmail.com, hc@pakistan.org.au

[3] CSIRO's Data61, Eveleigh, Sydney, Australia
didar.zowghi@csiro.au

## Abstract

As Artificial Intelligence (AI) transforms the domain of diplomacy in the 21st century, this research addresses the pressing need to evaluate the dualistic nature of these advancements, unpacking both the challenges they pose and the opportunities they offer. It has been almost a year since the launch of ChatGPT by OpenAI that revolutionised various work domains with its capabilities. The scope of application of these capabilities to diplomacy is yet to be fully explored or understood. Our research objective is to systematically examine the current discourse on Digital and AI Diplomacy, thus informing the development of a comprehensive framework for the role of Generative AI in modern diplomatic practices. Through the systematic analysis of 230 scholarly articles, we identified a spectrum of opportunities and challenges, culminating in a strategic framework that captures the multifaceted concepts for integration of Generative AI, setting a course for future research and innovation in diplomacy.

## Introduction

Artificial Intelligence (AI) has long established its relevance across a wide array of sectors. The Organisation for Economic Co-operation and Development (OECD)[1] defines Artificial Intelligence as "a machine-based system that, for explicit or implicit objectives, infers, from the input it receives, how to generate outputs such as predictions, content, recommendations, or decisions that can influence physical or virtual environments." With recent advancements, particularly in Generative AI solutions, the landscape of AI applications has broadened, especially due to its user-friendly, conversational interfaces and ability to help organisation in data and information driven decision making. Generative artificial intelligence (GenAI) is a branch of AI that can generate new content, be it text, images, audio, or video, which resembles human-generated contents[2]. Since the debut of OpenAI's ChatGPT in November 2022, there has been a surge of interest in analysing the impact of Generative AI across various fields of study and industries [1].

---

[1] https://oecd.ai/en/ai-principles

One domain experiencing a profound technological shift is diplomacy, where AI is becoming an indispensable asset for diplomats in performing their duties [2]. Diplomacy is a historically and culturally contingent bundle of practices that are analytically alike in their claim to represent a given polity to the outside world [3]. It can refer to established methods for influencing the decisions and behavior of foreign governments and peoples through dialogue, negotiation, and other measures. The landscape of diplomacy has undergone substantial changes in the last few decades, driven largely by the proliferation of the internet and the ubiquity of social media platforms [4-6]. These digital affordances have introduced new modalities for international relations, enabling rapid and direct communication across borders, fostering public engagement, and allowing for immediate dissemination and exchange of information. As a result, the traditional channels of diplomatic exchange have expanded beyond closed-door meetings and state-to-state dialogues to include digital interactions that engage a broader audience and operate in a more public and transparent manner [7, 8].

With these technological advancements, digital diplomacy has emerged as a pivotal tool, bringing both benefits and risks [9-11]. The benefits are manifold: increased efficiency, enhanced outreach, democratization of information, influencing public opinion, monitoring public sentiment, and the fostering of global dialogue. However, these come with inherent risks, such as misinformation, cyber threats, and the potential erosion of diplomatic protocols. The challenges of maintaining cybersecurity, ensuring accurate communication, and managing the fast pace of digital discourse are significant in this new era.

The advent of GenAI, with its advanced capabilities for content creation and data processing, stands to further revolutionize the field of diplomacy [12]. GenAI refers to AI models that can generate new content based on patterns learned from existing datasets. These models, such as OpenAI's GPT (Generative Pre-trained Transformer)[3] and DALL-E[4], Google's Bard[5] and Midjourney[6] have demonstrated remarkable capabilities in creating sophisticated outputs that often resemble those produced by humans. Their potential uses span across various domains, including healthcare [13], where AI can help personalize patient treatments, creative industries for generating art and literature, and automating tasks to enhance productivity. However, GenAI also faces limitations, including a propensity for bias, ethical concerns, and the need for large, high-quality datasets to train effectively [14]. Moreover, their outputs require careful scrutiny to ensure accuracy, appropriateness and integrity, as these systems may inadvertently generate false or misleading information, a phenomenon called 'hallucination'[15].

GenAI holds enormous potential to optimize the outcomes of diplomatic practices, thereby playing a pivotal role in crisis management, dispute resolution, and the promotion of trade, peace, security, and development [16]. In the rapidly changing diplomatic landscape, GenAI can automate and enhance analytical tasks, create simulations for conflict resolution scenarios, and even facilitate linguistic translations, all of which could potentially streamline diplomatic operations and open new avenues for engagement.

---

[2] https://cloud.google.com/use-cases/generative-ai
[3] https://openai.com/chatgpt
[4] https://openai.com/dall-e-3
[5] https://bard.google.com/
[6] https://www.midjourney.com/

The aim of this research is to conduct a thorough examination of the published literature on digital diplomacy and AI, to identify the predominant determinants and consequences of these technologies. By scrutinizing the evolving narrative and discourse through systematic and structured analysis, this study intends to propose a comprehensive framework that contextualizes the role of GenAI within modern diplomacy, offering insights and guidance for future diplomatic practices.

The contribution of this research lies in its comprehensive synthesis of existing literature and drawing insights from the personal multi-disciplinary experiences of the authors in Digital Transformation, AI and diplomacy, which provides a nuanced understanding of the intersection between GenAI and diplomacy. By examining over 230 scholarly articles, this study not only identifies the multifaceted opportunities and challenges of AI in this field but also critically assesses its potential impacts on diplomatic practices. The proposed framework, grounded in the insights gleaned from an extensive range of sources, represents a forward-looking contribution that aims to guide the integration of GenAI into the diplomatic sphere in a manner that is both innovative and cognizant of the delicate nature of international relations.

## Literature Review

The integration of cutting-edge technologies in diplomatic practices has garnered significant attention within foreign offices and across various domains such as bilateral, multilateral, cultural, public, development, and climate diplomacy. The surge in academic interest is reflected in our comprehensive search on Google Scholar and Scopus, which returned 230 unique entries, indicating a robust discourse on the merger of 'Diplomacy' with 'AI,' 'Artificial Intelligence,' or 'Digital'. Analyzing 230 abstracts and conclusions of these articles through thematic analysis has revealed a comprehensive map of digital and AI diplomacy. We acknowledge that our results are limited to the search terms and results retrieved from two sources. However, we consider 230 to be a good representative sample of research papers to provide us with required insights.

Despite the abundance of literature on AI and digital diplomacy, a noticeable gap exists—no articles specifically addressed 'ChatGPT,' 'LLM,' or 'Generative AI' within the context of diplomacy. The dearth of published literature on the use of GenAI in diplomacy can be attributed primarily to the novelty of the technology. GenAI is still an emerging field, and its implications, both beneficial and risky, are not fully understood within the complex and nuanced scope of international relations. Diplomacy, traditionally conservative and predicated on human judgment and discretion, necessitates a cautious approach when integrating new technologies that could fundamentally alter communication and negotiation processes. Furthermore, the potential for misuse of GenAI, such as the hallucinations [15, 17], creation of deepfakes [18], or the dissemination of misinformation [19], poses significant concerns that must be thoroughly evaluated before widespread adoption in diplomatic practices. This cautious approach is compounded by the need for rigorous ethical frameworks and international agreements on the use of such transformative technologies, which are still under development. There is also a gap in interdisciplinary expertise; the intersection of AI technology and diplomatic strategy is a highly specialized field that requires technologists and diplomats to bridge their knowledge divide. Hence, the integration of GenAI into diplomacy is not just a technological leap but also a significant policy and educational challenge that must be carefully navigated. Hence, under each category, when appropriate, we have provided our insights about GenAI, based on personal knowledge and experience when discussing the themes that emerged from the literature.

# Multifaceted domain of Digital and AI Diplomacy

Our analysis of the literature revealed several emergent themes, as will be described in this section.

1. **Digital Diplomacy [5, 7, 11, 20-135]**: This is the most prolific theme, capturing the transformative effects of digital tools and online platforms on international relations and statecraft. Digital diplomacy is the use of digital technologies and social media platforms to achieve foreign policy goals, enabling more direct communication and engagement with international and public audiences. Learning from literature, we can draw insights on how GenAI, with its ability to analyze vast data sets, predict trends, and generate human-like content, stands to significantly influence the diplomatic world. It can provide diplomats with enhanced tools for scenario planning, create sophisticated communication materials, and offer nuanced insights into public sentiment, potentially transforming the traditional methods of international relations and negotiation.

2. **Public Diplomacy [4, 6-8, 16, 47, 48, 66, 81, 83, 129, 134, 136-174]**: Extensive discussions in the literature can be found that has explored how AI and digital channels are reshaping the way states engage and communicate with public in a foreign state. Similarly, GenAI can revolutionize public diplomacy by providing tools to craft more targeted and resonant messaging, analyse public sentiment in real-time, and personalize engagement strategies across diverse cultural landscapes, thereby amplifying the reach and effectiveness of diplomatic initiatives.

3. **Health Diplomacy [13, 149, 175-183]:** The literature found explores the practice of using diplomatic negotiation and collaboration to address global health issues and advance health objectives in international policy, such as digital vaccine passports. Pandemic-related literature **[23, 33, 34, 57, 68, 81, 84, 86, 103, 118, 120, 154, 172, 176, 184, 185]** has explored the dimensions of digital diplomacy during the global pandemic of COVID-19. GenAI has the potential to significantly enhance health diplomacy by offering predictive analytics for public health crises, enabling data-driven policy decisions, and fostering cross-border cooperation on health technology innovations. It can also assist in tailoring health communication to diverse global audiences, ensuring that health initiatives and information are effectively disseminated and culturally sensitive.

4. **Foreign Policy [20-22, 55, 95, 97, 99, 135, 163, 167, 186]:** Exploring how AI can be utilized in foreign policy to enhance decision-making with predictive analytics, and tailor communication strategies to various international audiences. GenAI can significantly influence foreign policy by providing nuanced data analysis, predictive insights into international developments, and personalized diplomatic communication strategies, enhancing decision-making processes and enabling more agile, informed responses to global events.

5. **Negotiation [118, 187-196]:** Insights into how AI could impact diplomatic negotiations, potentially providing simulations and predictive models to inform strategies. GenAI, as exemplified by tools like Negotiator by ChatGPT, can enhance this process by simulating negotiations, predicting counterpart positions, and drafting diplomatic communications. Such AI tools offer opportunities for efficiency and data-informed strategies but also carry risks, including over-reliance on technology and potential misalignment with nuanced human judgment, underscoring the need for a balanced approach that combines AI assistance with skilled human diplomacy.

6. **Cultural Diplomacy [89, 92, 141, 166, 170, 185, 197-200]:** Discussions center on how AI can facilitate cultural exchanges, although concerns about cultural authenticity and representation are noted. GenAI can revolutionize this field by creating culturally resonant content, facilitating language translation for smoother intercultural communication, and personalizing engagement to align with diverse cultural values. However, it's essential to ensure that AI applications respect cultural nuances to avoid misrepresentation and maintain the authenticity of cultural exchanges.

7. **Social Media [4-8, 39, 57, 95, 122]:** There's a focus on social media's role in diplomacy, analyzing its use as a tool for public engagement, narrative dissemination, and soft power exertion. There are many examples of social media use in diplomacy. **X (formerly known as Twitter)[7, 46, 49, 63, 66, 86, 92, 101, 109, 120, 121, 137, 139, 153, 197, 201]** Twitter is used in diplomacy as a tool for public engagement, allowing diplomats and state leaders to communicate directly with international audiences, shape narratives, and respond swiftly to global events. **Facebook [63, 66, 67, 102, 133]** Facebook is employed in diplomacy for its vast networking capabilities, enabling diplomatic entities to engage with the public, disseminate information, and foster digital dialogues with global stakeholders. **Bots [177, 199, 202]** Diplomatic bots on social media platforms are used to automate routine communications, disseminate information, and facilitate outreach, effectively scaling up diplomatic engagement efforts. GenAI can enhance social media's diplomatic use by creating nuanced content and analyzing public sentiment, allowing for more strategic communication. However, risks include the potential for spreading misinformation, the amplification of biases, and the reduction of complex diplomatic stances into oversimplified messages that might not convey the full context or intentions of a nation's foreign policy.

8. **Security Diplomacy [179, 203-207]:** Exploration of AI's applications in security-related diplomatic endeavors, such as threat analysis and peacekeeping operations. GenAI can significantly impact this domain by analyzing security threats, enhancing cybersecurity measures, and simulating conflict scenarios for better strategic planning. However, the reliance on GenAI for security decisions carries the risk of escalating conflicts if not managed with caution and deep human oversight due to the complexity and sensitivity of security-related communications.

9. **Economic Diplomacy [184, 208-210]:** A smaller focus on the economic implications of AI in diplomacy, particularly in trade and international economic policy. GenAI stands to revolutionize this arena by providing predictive analytics for market trends, automating the analysis of economic agreements, and simulating trade scenarios. It enables diplomats to negotiate from a data-enriched position, potentially leading to more favourable outcomes. However, care must be taken to ensure that AI's predictive models are free from biases that could misguide economic policies or create unfair advantages, maintaining the integrity of economic diplomacy.

10. **Environmental Diplomacy [192, 211, 212]:** Limited but crucial examinations has been provided in literature on the AI's role in addressing global environmental challenges through diplomatic channels. GenAI can influence this field by providing sophisticated models for climate prediction, analyzing environmental data to inform policy decisions, and simulating the outcomes of environmental agreements. However, ensuring the accuracy of AI-generated environmental data is crucial, as misinterpretations could lead to ineffective or detrimental policies, underscoring the need for meticulous validation against empirical data.

11. **Risk Assessment [93, 213]:** Not much focus has been given to specific analyses of the risks and uncertainties posed by the integration of AI into diplomatic practices. Risk assessment for the use of GenAI in diplomacy must address the potential for algorithmic bias, which can skew diplomatic communications and negotiations. The misinterpretation of generated data and the possibility of manipulative information campaigns are also concerns, particularly in sensitive geopolitical contexts. Additionally, the reliance on AI for decision-making processes could lead to an overestimation of AI capabilities, creating vulnerabilities in diplomatic strategies. Hence, continuous monitoring, transparent review processes, and the incorporation of human oversight are critical to mitigating these risks and ensuring AI's constructive role in diplomacy.

12. **Conflict Resolution [39]:** one article has explored AI's contribution to conflict resolution. GenAI can offer novel opportunities in conflict resolution by simulating complex negotiation scenarios and generating creative solutions that might elude human diplomats. Its ability to process vast amounts of historical and real-time data can inform more strategic and effective resolutions. In terms of risks; AI may oversimplify the nuances of human conflict or encode biases present in the data it learns from, potentially leading to incomplete or one-sided solutions. Ensuring AI-generated proposals are critically evaluated by experienced diplomats is crucial to harnessing its benefits while mitigating its limitations.

13. **Challenges [9-11, 59, 85, 100, 185, 187, 198, 206, 207, 214, 215]:** Significant attention is given to the hurdles presented by AI, including ethical considerations, the digital divide, and potential geopolitical tensions arising from uneven AI adoption. GenAI can help overcome some of these challenges by providing advanced tools for ethical decision-making, bridging the digital divide through accessible AI education, and offering simulation models to anticipate and manage geopolitical tensions. Additionally, it can facilitate a more equitable distribution of AI benefits by generating customized solutions that address the specific needs of diverse global populations. There are numerous existing resources on responsible and ethical AI [216] as well as diversity and inclusion in AI [217, 218] that could guide the integration of GenAI in diplomacy.

14. **Opportunities [9-11, 100, 177, 182, 185, 200]**: Researchers delve into the potential of AI to enhance diplomatic processes, from data-driven decision-making to virtual engagements and simulations.

15. **Case Studies [24, 64, 156, 164, 195]:** There is a moderate emphasis in literature on real-world applications and the tangible impact of AI and digital technologies in diplomatic contexts.

16. **Disruptive Technologies in Diplomacy [95, 183, 209, 219-221]:** Some articles have discussed the potential of new technologies to disrupt established diplomatic protocols and procedures.

## Trend over the years

Figure 1 presents a time series analysis of the number of publications from 1988 to approximately 2023. The graph shows a relatively stable and low number of publications until the early 2010s, when there was a noticeable increase. In the early literature exploring the intersection of AI or digital technologies, and diplomatic processes, papers from as early as 1988 began to surface. For example, Kraus and Lehmann [212] describe an AI program capable of playing the board game Diplomacy. This paper, along with others like [192] by the same authors, and Sangineto's work [194] on multi-agent negotiation, indicates an initial focus on strategic negotiation and decision-making within defined rule-based systems. The usage of diplomacy in these contexts aligns more with strategic gameplay rather than geopolitical diplomacy. However, these early explorations laid the groundwork for understanding complex interactions and negotiations, foundational elements of diplomatic discourse, even if the term 'diplomacy' was applied in a more specialized context. The most striking insight is a sharp rise in publications after 2015, reaching a peak around 2020. This trend indicates a growing academic and professional interest in the intersection of diplomacy with digital and AI technologies, likely reflecting the increasing importance of these technologies in global diplomatic practices.

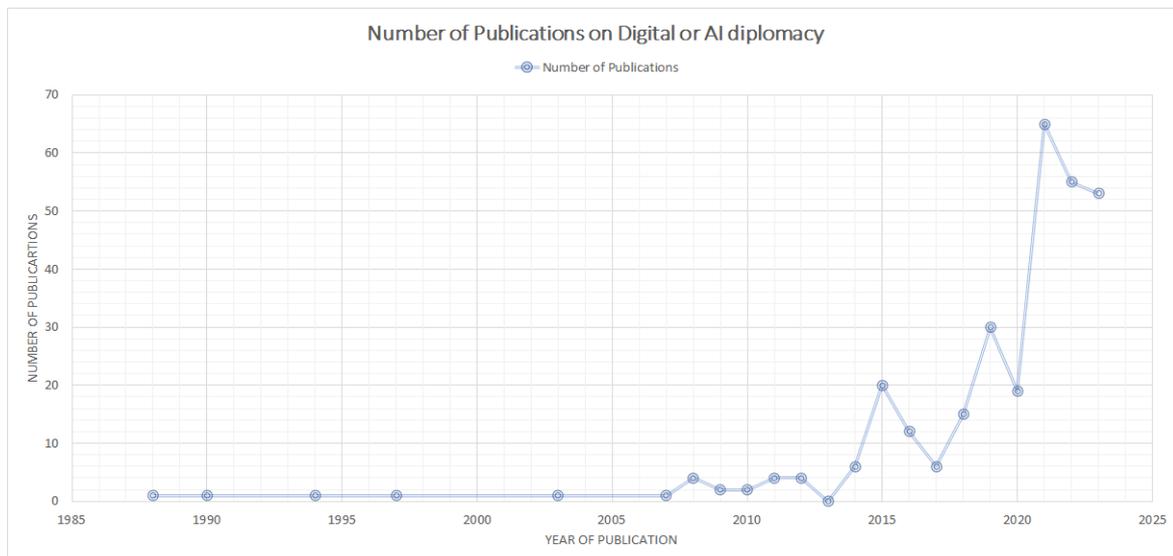

Figure 1. Trends over the years

## The geo-political focus of the studies

The global landscape of AI in diplomacy and digital diplomacy is marked by varying levels of engagement or focus across different countries, continents, and regions.

- Afghanistan: [127]
- Africa: [30, 51, 57, 69, 82, 94, 117, 124, 133, 159, 222-225]
- Asia: [55, 119, 149, 226]
- Australia: [8]
- Canada: [43, 163, 205]
- China: [6, 88, 94, 104, 121, 123, 125, 132, 133, 149, 154, 155, 159, 172, 227, 228]
- Europe: [6, 59, 86, 122, 128, 229-232]
- Egypt: [24]
- Germany, Italy, Spain: [226]
- India: [55, 68, 158, 162, 233]
- Indonesia: [113, 114, 184, 209]
- Iran: [37, 123, 173]
- Ireland: [205]
- Israel: [102, 140]
- Japan: [6, 71, 90, 200]

- Kazakhstan: [67, 112]
- Kenya: [205]
- Middle East: [30, 135, 136, 138]
- Pakistan: [130]
- Russia: [70, 104, 110, 115, 122, 201, 210, 234]
- South Korea: [71, 85, 90, 166, 176]
- Sweden: [22, 27, 97, 153]
- Taiwan: [87]
- Ukraine: [70, 104, 174, 201]
- United Kingdom: [24, 43, 88]
- USA: [63, 88, 94, 99, 101, 156, 173]

The geographical spread of research on AI and digital diplomacy, as indicated by the literature, suggests an increasing academic interest, though this does not necessarily mirror the actual global adoption and implementation. The high volume of publications about China, alongside emerging research in Africa and consistent contributions from Europe and India, signals a keen academic pursuit in understanding the potential of AI in diplomacy. However, this trend, with its sporadic distribution from regions like Australia, the Middle East, and the UK, should be interpreted with caution, as the literature may not fully represent the on-the-ground reality, being limited to the scope of academic publications.

In the global context of diplomacy, GenAI presents opportunities to streamline negotiations, craft nuanced communication strategies, and analyze complex international relations data. Countries at the forefront of AI research, such as the United States and China, may leverage these advancements to gain a diplomatic edge, potentially exacerbating global power imbalances. To ensure a level playing field, inclusive practices and international cooperation are vital in sharing AI advancements and fostering a multilateral approach to technology governance. This could mitigate risks such as the monopolization of AI benefits and ensure that all nations can harness the potential of AI in diplomacy. The UN sees GenAI as a potential tool to advance its Sustainable Development Goals (SDGs), recognizing its capacity to generate new content or ideas that could contribute to solving complex global issues[7]. The UN's guidance on GenAI advocates for a human-centred approach, emphasizing the protection of core humanistic values like human agency, inclusion, equity, gender equality, and linguistic and cultural diversities [235]. It acknowledges the risks posed by GenAI and proposes regulatory steps to safeguard data privacy and ethical usage[8]. The guidance also emphasizes the need for educational institutions to validate GenAI systems for their ethical and pedagogical appropriateness and urges international reflection on the long-term implications of GenAI for knowledge and learning.

## Opportunities and Benefits

The convergence of AI with diplomatic practices creates new opportunities as outlined by contemporary scholarship. Roumate [11] acknowledges AI's role in enhancing international collaboration and amplifying the influence of new actors in global governance, while also addressing the potential for AI to facilitate international regulation and the interplay of AI with cyber diplomacy. Roumate further highlights the utility of social media and AI as tools for digital diplomacy and the critical aspects of cybersecurity and data sovereignty. Grincheva [185] contributes to this discourse by emphasizing the acceleration of digital cultural production, particularly in the face of global crises, and the complementing of official cultural diplomacy through digital initiatives. Further expanding the scope, Bansal, Kunaprayoon, and Zhang [177] advocate for robotic telesurgery's role in global health diplomacy, and Konovalova [10] underscores the importance of creating stability in cyberspace and enhancing the diplomatic capacity with AI to achieve national objectives effectively.

___________________________________________________________________________________________

[7] https://webtv.un.org/en/asset/k14/k14ar7aqzw
[8] https://www.sciencediplomacy.org/perspective/2022/united-nations-focal-diplomatic-role-emerging-technologies

The opportunities for AI in diplomacy, are expansive and transformative [9-11]. AI's potential to bolster international cooperation, enhance the quality of diplomatic training, predict political crises, and anticipate humanitarian catastrophes is underscored. In [10] Konovalova points to AI's role in improving encryption and surveillance through quantum computing. As an instrument of traditional and public diplomacy, AI is posited as a significant player in the geopolitical landscape, influencing the balance of power. These developments are leading to the creation of virtual embassies, the use of bots for citizen communication, and the appointment of digital ambassadors, signalling a shift toward a more dynamic and technologically integrated approach to diplomacy. This evolution in diplomacy is aligned with the broader aim of achieving global goals, including democracy, peacebuilding, human rights, and environmental protection.

Significant advancements in technology, particularly AI, are offering novel ways to augment the capabilities of diplomatic corps [10]. A noticeable example is a Hungarian innovation—a blockchain-based digital application tailored for cyber diplomacy—that enhances traditional diplomatic functions [236]. This application augments communication security through encryption and stratified access, catering to the global diplomatic community. It specifically bolsters cyber diplomacy by managing the multifaceted implications of emergent digital technologies, ensuring that the diplomatic sector capably navigates the evolving landscape of international relations.

In [9], the authors have discussed how AI presents a transformative opportunity for diplomacy, particularly during crises such as conflicts, migrations, and pandemics. It offers advanced tools for monitoring, analyzing, and crafting responses, leveraging machine learning for a range of tasks including classification, generation, and prediction. Research has shown AI's utility in peacebuilding and security negotiations at the UN and in strategic reasoning for intervention planning. AI-assisted tools enhance consular services and resource allocation, manage public expectations, and facilitate communication. The integration of non-traditional data, like social media and satellite imagery, into situational awareness tasks, further exemplifies AI's potential. However, the development of AI in diplomacy necessitates responsible data sharing, interpretable and robust AI models, and capacity building among crisis response personnel to fully harness these emerging technologies.

## Challenges and Risks

AI is an umbrella term that encompasses broad range of technologies. AI itself has limitation in terms of quality and currency of data, biases, fairness, manipulations, and black box nature [217]. These existing issues and additional challenges need further consideration and risk assessment before integrating AI in diplomatic processes. For example as indicated by [9] for use of AI-enabled decision making during crisis events, firstly, the inherent limitation of incomplete information during crises, compounded by deliberate data obfuscation, demands AI tools adept at handling and visualizing uncertainties. Secondly, the high stakes of human life necessitate AI outputs to be explainable, transparent, and secure. Thirdly, diplomatic decision-making, deeply influenced by cultural nuances and value judgments, requires AI that can integrate domain knowledge and experience. Lastly, the multi-layered nature of diplomatic interests during crises often leads to mistrust and strategic misuse of information, with social media data being particularly vulnerable to misinformation campaigns.

The deployment of AI in diplomacy introduces several risks and challenges [10]. The advent of AI may disrupt the traditional role of diplomats as machines begin to process vast quantities of data, potentially overshadowing human judgment. The speed at which AI and machine learning can report and disseminate information demands a re-evaluation of diplomatic operations, particularly the autonomy of diplomatic missions and their real-time engagement with social media. Moreover, AI's influence on public diplomacy is double-edged; while it can optimize resource allocation and strategic planning, it also raises concerns about misinformation, the erosion of diplomatic empathy, and the potential loss of nuanced human judgment in complex, value-laden decisions. This paradigm shift necessitates a reimagining of diplomatic roles and a reaffirmation of essential human skills within AI-enhanced diplomacy.

AI's inherent biases and issues with fairness [217], pose considerable risks for its application in diplomacy, where trust and accuracy are paramount. Given that diplomatic decision-making and negotiations rely heavily on impartiality and the nuanced understanding of complex human factors, the propensity of AI to reflect the prejudices present in its training data can lead to skewed outcomes. Such limitations can undermine the credibility of AI-assisted diplomatic efforts and potentially jeopardize sensitive international relations.

Table 1 reflects a spectrum of challenges and risks associated with the integration of AI across various domains of diplomacy. It highlights concerns such as the misalignment with national interests in Bilateral Diplomacy, coordination complexities in Multilateral Diplomacy, and the potential for cultural misrepresentation in Cultural Diplomacy. Public Diplomacy faces risks of misinformation, while Security Diplomacy could suffer from an over-reliance on AI. Data inaccuracies in Environmental Diplomacy, ethical dilemmas in Humanitarian Diplomacy, and cybersecurity threats in Digital Diplomacy represent other significant challenges. In Conflict Resolution, the lack of human empathy in AI-generated solutions is a notable concern. Inequitable access to AI can put diplomats from technologically less advanced countries in a disadvantaged position, further eroding trust in complex scenarios.

| **Domain of Diplomacy** | **Challenges and Risks of integrating AI in Diplomacy** |
|---|---|
| Bilateral Diplomacy | Misalignment with national interests, sensitive information leaks, inequitable access to AI |
| Multilateral Diplomacy | Complexity in coordination, data privacy concerns across borders |
| Economic Diplomacy | AI biases affecting economic agreements, misinterpretation of economic data |
| Cultural Diplomacy | Loss of cultural nuances, potential for cultural misrepresentation |
| Public Diplomacy | Misinformation, challenges in audience engagement across diverse cultures |
| Security Diplomacy | Over-reliance on AI for security decisions, potential for escalation |
| Environmental Diplomacy | Data inaccuracies leading to ineffective environmental policies |
| Humanitarian Diplomacy | Ethical dilemmas in AI decision-making during crises |
| Digital Diplomacy | Cybersecurity risks, ethical challenges in digital engagement |
| Conflict Resolution and Mediation | AI-generated solutions may lack human empathy and understanding |

Table 1. Potential Challenges and Risks of integrating AI in various domains of Diplomacy.

While AI in diplomacy opens a new wave of opportunities, enhancing efficiency and creating new avenues for engagement, it is not without its challenges and risks. Biases, data privacy concerns, and the potential erosion of trust necessitate a cautious approach. Therefore, there is a pressing need for further exploration and research to develop robust frameworks that ensure AI's ethical and effective integration into diplomatic practices, fortifying trust and accuracy in high-stakes international relations.

# Framework for Generative AI in Diplomacy

Our thorough literature review, along with our extensive experiences in the domains of AI and diplomacy, have provided us with a broader perspective on the multifaceted landscape of AI in diplomacy. We were able to identify opportunities where GenAI could augment diplomatic efforts, from automating routine tasks to enhancing the outcomes of strategic negotiations. Simultaneously, it has underscored several outstanding challenges that accompany the adoption of AI, such as ethical dilemmas, the potential for misuse, and the need for robust cyber-security measures. These insights have provided us with a foundation for exploring themes to construct a framework for integrating GenAI in diplomacy. Despite the comprehensive coverage, current research falls short in reviewing the implications of GenAI, specifically within the research area of diplomacy. To address this gap, based on our extensive review of literature and personal experiences, we propose a framework for GenAI across various forms of diplomacy (as shown in Figure 2) are described below:

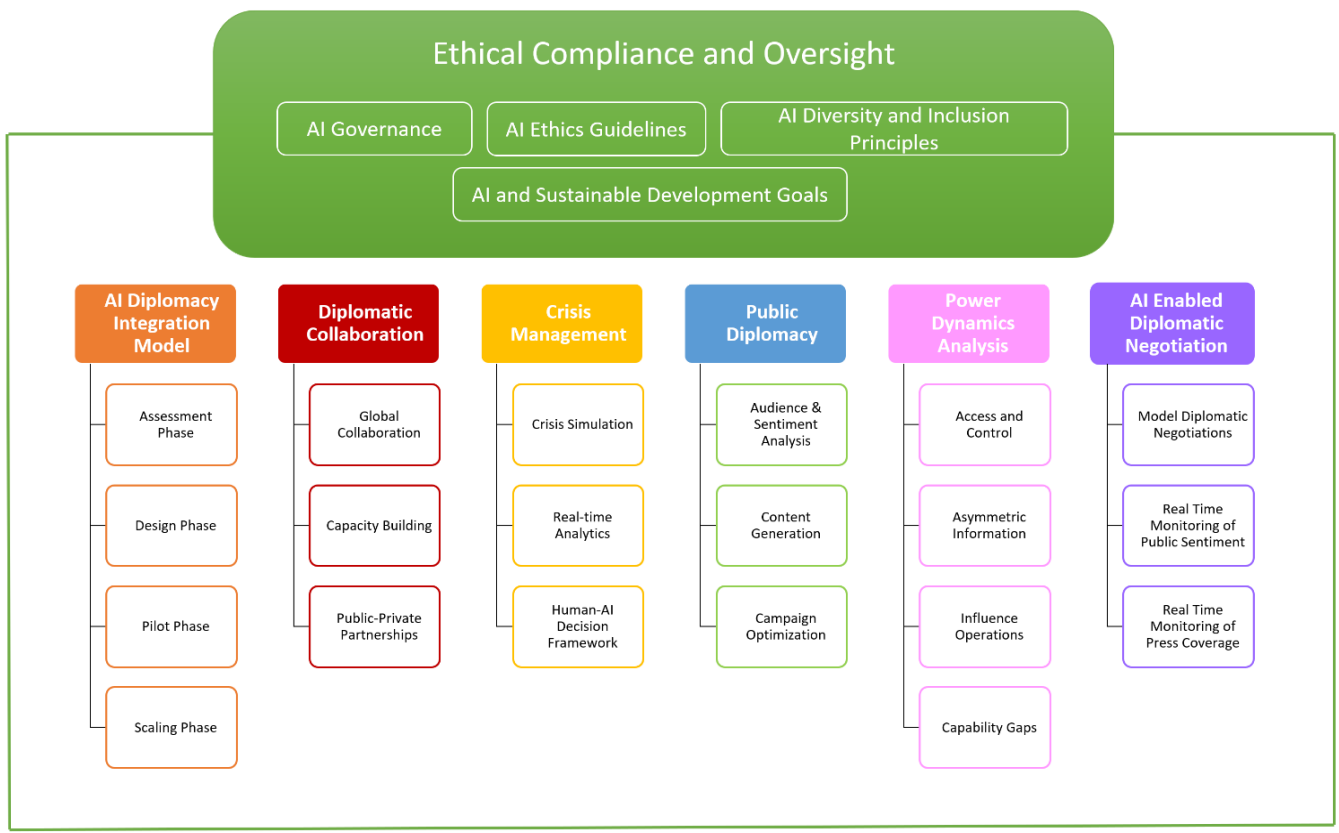

Figure 2. Framework for GenAI in Diplomacy

There are five independent and intertwined components of the proposed framework: AI Diplomacy integration model, Diplomatic Collaboration, Crisis Management, Public Diplomacy and Power Dynamic Analysis. All components must adhere to the Ethical Compliance and Oversight, which provides resources for AI Governance, Ethics Guidelines and Diversity and Inclusion Principles.

## Ethical Compliance and Oversight

This layer encompasses all the components of the framework to ensure that the application of GenAI in diplomacy, adheres to moral principles, legal standards, equity, and inclusive practices. The necessity of Ethical Compliance and Oversight in the application of GenAI in diplomacy is underscored firstly by the imperative of ensuring equitable access to AI technologies, especially in supporting developing countries. This helps in bridging the digital divide and fostering a balanced global diplomatic landscape. Secondly, the active involvement of the diplomatic community in dialogues on responsible AI governance is crucial. Their participation ensures that international collaboration in AI aligns with the intricate nuances and ethical considerations unique to diplomatic relations, promoting universally beneficial and responsible use of AI in international affairs.

- AI Governance [237]: To ensure responsible design, development, and use of GenAI; in diplomacy, it sets the legal and operational framework for AI's role in diplomacy.
- AI Ethics Guidelines [238]: Offer a moral compass for AI deployment; in diplomatic contexts, they help navigate ethical dilemmas ensuring AI respects international cultural norms.
- AI Diversity and Inclusion Principles [217]: Aim to promote equity and representation in AI systems; within diplomacy, they ensure that AI tools and their outputs do not perpetuate biases and reflect the diverse global stakeholders.
- AI and Sustainable Development Goals [239]: Aim to align GenAI applications with the Sustainable Development Goals to reinforce global diplomatic efforts toward sustainable development and equitable progress.

Various nations and organisations worldwide have proposed AI ethics and governance frameworks. China's Beijing AI Principles[9] emphasize research, development, and long-term planning, while NIST in the United States has established principles for explainable AI, underscoring the need for transparency and accountability [240]. These efforts, alongside Australia's AI Ethics Principles[10], the European Union's Ethics Guidelines for Trustworthy AI[11] and AI Act[12], UNESCO's global standards[13], World Economic Forums' 9 ethics principles[14], and the OECD's stewardship principles[15], constitute a global attempt to ensure AI's alignment with ethical norms, human rights, and democratic values. As GenAI continues to evolve, the adoption of ethical frameworks becomes increasingly vital. These frameworks are critical for ensuring that AI advances respect moral, legal, and cultural norms, particularly within diplomacy. The principles of AI governance, ethics, and diversity must be at the forefront of this technological frontier to foster an inclusive and unbiased global diplomatic landscape and to bridge the digital divide.

---

[9] https://perma.cc/V9FL-H6J7
[10] https://www.industry.gov.au/publications/australias-artificial-intelligence-ethics-framework/australias-ai-ethics-principles
[11] https://digital-strategy.ec.europa.eu/en/library/ethics-guidelines-trustworthy-ai
[12] https://www.europarl.europa.eu/news/en/headlines/society/20230601STO93804/eu-ai-act-first-regulation-on-artificial-intelligence
[13] https://www.unesco.org/en/artificial-intelligence/recommendation-ethics
[14] https://www.weforum.org/agenda/2021/06/ethical-principles-for-ai/
[15] https://oecd.ai/en/ai-principles

## AI Diplomacy Integration Model

This component aims to ensure a structured approach for integrating AI into diplomatic practices.

- Assessment Phase: Evaluate current diplomatic processes to identify potential areas for AI integration.
- Design Phase: Design and develop AI tools tailored to specific diplomatic functions, ensuring they align with ethical standards and international norms.
- Pilot Phase: Implement AI on a small scale within selected diplomatic domains to gather data and refine the technology, including testing, verification and validation.
- Scaling Phase: Gradually expand the use of AI tools in diplomacy based on the insights from the pilot phase, including monitoring and assessment.

## Diplomatic AI Collaboration

This component of the framework aims to foster global cooperation in the realm of AI and diplomacy.

- Global Collaboration: Foster international collaboration for sharing best practices and AI resources.
- Capacity Building: Provide training for diplomats and technical staff to effectively use AI tools.
- Equitable Access to AI: Advocacy for equitable access to AI technologies for all nations, with particular focus on supporting developing countries.
- Public-Private Partnerships: Engage with tech companies and academia to advance AI research relevant to diplomacy.

## AI Enabled Diplomatic Negotiations

This component is aimed at enhancing the outcomes of diplomatic negotiations, both bilateral as well as multilateral, through use of AI technologies, including GenAI.

- Model Diplomatic Negotiations: Model diplomatic negotiations can comprise two sets of negotiations, one with and the other without the help of AI tools. The outcomes of two sets of negotiations can be used to analyse the impact of applications of AI tools on the outcomes of diplomatic negotiations.
- Real-Time Monitoring of Public Sentiment and Press Coverage: AI tools can be used to monitor public sentiment and press coverage of the ongoing diplomatic negotiations. The impact of such monitoring on the respective positions of negating sides and the outcome of diplomatic negotiations can be analysed.

## AI Crisis Management

This is a strategic component aimed at enhancing the capacity of diplomatic entities to manage crises through AI.

- Crisis Simulation: Use AI to simulate diplomatic crises for better preparedness.
- Real-time Analytics: Develop AI systems capable of providing real-time analytics during crises.

- Human-AI Decision Framework: Define clear protocols for human oversight in AI-assisted decision-making during crises.

## Public Diplomacy

This component of the framework focuses on leveraging AI to understand and engage international audiences effectively.

- Audience and Sentiment Analysis: Utilize AI to understand and segment audiences for targeted public diplomacy and conduct sentiment analysis.
- Content Generation: Employ GenAI to create personalized content that resonates with different cultures and demographics.
- Campaign Optimization: Use GenAI to optimize public diplomacy campaigns for greater impact and engagement.

## Power Dynamics Analysis

This component examines how AI influences the balance of power in diplomacy and international relations.

- Access and Control: Analyse how disparities in access to AI capabilities can affect diplomatic influence and negotiation power.
- Asymmetric Information: Examine the impact of GenAI on creating or reducing information asymmetries in international relations.
- Influence Operations: Assess the use of AI in shaping public opinion and its effectiveness in psychological operations within the diplomatic sphere.
- Capability Gaps: Identify strategies to address the divide between nations with varying levels of AI technology, aiming to prevent further imbalances in diplomatic interactions.

This framework aims to integrate GenAI into diplomacy in a way that enhances diplomatic processes while addressing potential ethical concerns and fostering international cooperation.

# Illustrative Case Studies

These examples illustrate how AI can significantly contribute to the efficiency and effectiveness of diplomatic services. GenAI holds the promise of further enhancing these capabilities but also poses risks that require careful consideration and management.

## Example Case 1: GenAI in Consular Services

> In consular services, GenAI can be used to manage high demand for emergency passports, visa requests, and business certifications. AI can assist in identifying patterns in application demands, leading to more effective management of consular requests. GenAI could further improve this by generating predictive models for demand, reducing wait times, and increasing efficiency.
>
> Risks: If GenAI produces inaccurate predictions, it could worsen backlogs and reduce public trust.

## Example Case 2: GenAI in Crisis Management

GenAI systems can be valuable in helping diplomats manage crises by analyzing real-time data to understand the nature and gravity of events.

Risk: When faced with unanticipated situations, potentially exacerbating a crisis if not carefully managed due to internal limitations of AI models.

## Example Case 3: GenAI in Public Diplomacy

GenAI can be used in a public diplomacy campaign when there is a limited budget. AI can analyse social media data to gauge public interest and reception in various policy priorities, helping to decide on the most effective campaign strategy. GenAI can help in creating more engaging content for the campaign or simulating different public reactions.

Risk: If GenAI misinterprets public sentiment or cultural sensitivity, it could lead to ineffective campaigns and wasted resources, and worst backfire.

## Example Case 4: GenAI in Diplomatic Negotiations

GenAI can be used in diplomatic negotiations, both bilateral as well as multilateral. It can be a powerful tool to optimize the outcomes of diplomatic negotiations by quickly providing and analysing information that can be critical to the negotiations. AI can also create different scenarios for the negotiating sides better informing them about the possible implications of various outcomes of diplomatic negotiations.

Risk: confidentiality of diplomatic negations and respective positions of negotiating sides can be compromised.

# Conclusion and Future Work

Our extensive literature review of 230 articles on AI and digital diplomacy has illuminated the diverse landscape of AI's applications across various diplomatic domains. We have underscored both the potential benefits and the inherent challenges, offering a balanced perspective on this emergent technology's role in statecraft. Our proposed framework for employing GenAI in diplomacy seeks to navigate these complexities, aiming to leverage AI's capabilities while mitigating its risks.

For future work, we advocate for empirical research to test and refine the proposed framework within real-world diplomatic contexts. There is a significant opportunity to conduct case studies and pilot programs with diplomatic institutions to evaluate the practical implications of GenAI. Additionally, interdisciplinary research involving AI ethics, international law, and diplomatic studies will be crucial to developing comprehensive guidelines and codes for the responsible use of AI in diplomacy. This future work should strive to align AI advancements with the fundamental principles of diplomacy—trust, mutual respect, and the peaceful resolution of conflicts.

# Disclosure Statement

No potential conflict of interest was reported by the authors.

# Notes on Contributors

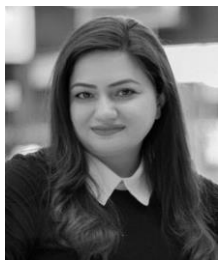

**1st Author: Muneera Bano, PhD** is Senior Research Scientist and member of the AI Diversity and Inclusion team at CSIRO's Data61. She is an award-winning scholar and is a passionate advocate for gender equity in STEM. She is a Diversity Inclusion and Belongingness (DIB) officer at Data61 and a member of the 'Equity, Diversity and Inclusion committee for Science and Technology Australia. Muneera graduated with a PhD in Software Engineering from UTS in 2015. She was granted the Dr. John Yu Fellowship for 'Cultural Diversity and Leadership' at Sydney University in 2019, and the 'Pathways to Politics leadership program by Melbourne University in 2021. She was awarded the title of Superstar of STEM and 'Most Influential Asian-Australian Under 40,' in 2019, and she was honoured by the Government of Pakistan as an under-40 leader in science and innovation, including in the 'Foreign Minister's Honour List' in 2021. She has published more than 50 research articles in notable international forums on Software Engineering. Her research, influenced by her interest in AI and Diversity and Inclusion, emphasizes human-centric technologies.

Contact her at muneera.bano@csiro.au
ORCID: https://orcid.org/0000-0002-1447-9521
Official Webpage: https://people.csiro.au/B/M/muneera-bano

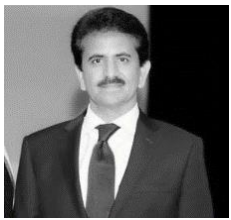

**2nd Author: Ambassador Zahid Hafeez Chaudhri** is a senior diplomat with over 30 years of experience in bilateral and multilateral diplomacy. He is currently serving as High Commissioner for Pakistan to the Commonwealth of Australia, Fiji, Papua New Guinea, Solomon Islands, Vanuatu, and Nauru. Previously he has served at key Pakistan Missions, including the Embassy of Pakistan, Washington DC; High Commission for Pakistan, London; Consulate General of Pakistan, Dubai; and Embassy of Pakistan, Paris. He has also served as Additional Foreign Secretary of Pakistan for Asia Pacific and Spokesperson of the Ministry of Foreign Affairs, Islamabad; Joint Secretary National Security at the National Security Division of Pakistan; and Director General for Economic Coordination at the President's Secretariat. Ambassador Chaudhri has also represented Pakistan on the Commonwealth Secretariat Board of Governors for 3 years. He has produced and presented several research & professional reports and policy papers for the Government and other key stakeholders. He is passionate about the use of AI in diplomatic practices. Ambassador Chaudhri holds a Master's degree in International Law from the University of London and second Master's in Business Administration. He is also an alumnus of the National Defence University, Washington DC. He is a British Chevening scholar. Contact him at hc@pakistan.org.au

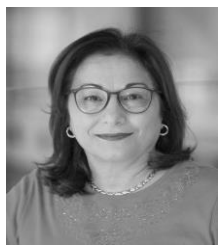

**3rd Author: Didar Zowghi,** (PhD, IEEE Member since 1995) is a Senior Principal Research Scientist and leads the science team for Diversity and Inclusion in AI at CSIRO's Data61. She is an Emeritus Professor at the University of Technology Sydney (UTS) and conjoint professor at the University of New South Wales (UNSW). She has decades of experience in Software Engineering research and practice. In 2019 she received the IEEE Lifetime Service Award for her contributions to the research community, and in 2022 the Distinguished Educator Award from IEEE Computer Society TCSE. She has published over 220 research articles in prestigious conferences and journals and has co-authored papers with over 100 researchers from 30+ countries.

Contact her at didar.zowghi@csiro.au
ORCID: https://orcid.org/0000-0002-6051-0155
Official Webpage: https://people.csiro.au/z/D/Didar-Zowghi